\documentclass[5p]{elsarticle}
\usepackage{hyperref}
\usepackage{amssymb,amsmath}
\usepackage{lineno}
\usepackage{natbib}
\usepackage{wrapfig}
\usepackage{color}
\usepackage{url}
\usepackage{multirow}
\usepackage{epstopdf}
\usepackage{soul}
\usepackage{graphicx}
\usepackage[dvips]{epsfig}
\usepackage{subfig}
\usepackage{caption}
\usepackage{chngcntr}
\usepackage{geometry}
\usepackage[export]{adjustbox}
\usepackage[utf8]{inputenc} 
\usepackage{amsfonts}       
\usepackage{nicefrac} 
\usepackage{gensymb} 
\usepackage{enumitem}

\hypersetup{colorlinks=true,
			linkcolor=black, 
			citecolor=blue, 
			urlcolor=blue, 
			pdfborder={0 0 0}}

\journal{Nuclear Instruments and Methods in Physics Research A}

\newcommand{\comment}[1]{}
\journal{Nuclear Instruments and Methods in Physics Research A}
\modulolinenumbers[1]

\begin{document}

\begin{frontmatter}
\title{Development of a multi-channel power supply for silicon photo-multipliers used with inorganic scintillators}
\author{O. Javakhishvili$^{1,2}$, I. Keshelashvili$^1$, D. Mchedlishvili$^{3,4}$, M. Gagoshidze$^2$, T. Hahnraths$^1$, A. Kacharava$^1$,\\ Z. Metreveli$^2$, F. M{\"u}ller$^1$, T. Sefzick$^1$, D. Shergelashvili$^4$, H. Soltner$^5$, and H. Str{\"o}her$^1$
for the JEDI collaboration.
} 
%%%%%%%%%%%%%%%%%%%%%%%%%%%%%%%%%%%%%%%%%%%%%%%%%%%%%%%%%%%%%%%%%%%%%%%%%%%%%%%%%%%
\address{
$^1$ Institute of Nuclear Physics (IKP), Forschungszentrum J{\"u}lich, 52428 J{\"u}lich, Germany\\
$^2$ Department of Electrical and Computer Engineering, Agricultural University of Georgia, 0159 Tbilisi, Georgia\\
$^3$ High-Energy Physics Institute, Tbilisi State University, 0186 Tbilisi, Georgia \\
$^4$ Tbilisi State University, SMART$|$EDM Laboratory, 0179 Tbilisi, Georgia \\ 
$^5$ Institute of Engineering and Technology (ZEA-1), Forschungszentrum J{\"u}lich, 52428 J{\"u}lich, Germany \\}


\begin{abstract}
The motivation of the current R$\&$D project is based upon the requirements of the JEDI international collaboration \footnotemark \fntext[JEDI]{\url{http://collaborations.fz-juelich.de/ikp/jedi/}} aiming to measure Electric Dipole Moments (EDMs) of charged particles in storage rings.  One of the most important elements of such an experiment will be a specially designed polarimeter with the detection system  based on a modular inorganic scintillator (LYSO crystal) calorimeter. The calorimeter modules are read out by Silicon Photo Multipliers (SiPMs). This paper describes the development of a multi-channel power supply for the polarimeter modules, providing very stable and clean bias voltages for SiPMs. In order to ensure the best possible performance of SiPMs in conjunction with the crystal-based calorimeter modules and to guarantee the required level of calorimeter stability, several quality requirements have to be met by the power supply. Additionally, it is required to provide features including remote control via the network, ramping of the output voltage, measuring and sending the information about its output voltages and currents, etc. 
The obtained results demonstrate that the goals for the JEDI polarimeter are met. The developed hardware will be useful in other fields of fundamental and applied research, medical diagnostic techniques and industry, where SiPMs are used.
\end{abstract}
%%%%%%%%%%%%%%%%%%%%%%%%%%%%%%%%%%%%%%%%%%%%%%%%%%%%%%%%%%%%%%%%%%%%%%%%%%%%%%%%%%%
\begin{keyword}
Precision measurements \sep LYSO scintillating material \sep hadron calorimeter
\sep light sensors 
\end{keyword}
\end{frontmatter}

\section{Introduction \label{sec:introduction}}
The observed Matter-Antimatter asymmetry in the Universe cannot be
explained by the Standard Model (SM) of Particle Physics.  In order to resolve the mystery of matter dominance an additional {\it CP}-violating phenomenon is needed. A crucial aspect in the search for physics beyond the SM will be a non-vanishing Electric Dipole Moment (EDM) of subatomic particles \cite{art:sakharov}.

The interaction of a particle's spin with strong electric fields enables the measurement of an EDM.
The envisaged experiments need to be performed with high-precision storage rings and require
an accurate measurement and control of the spin and the beam.

\begin{figure}[ht]
  \includegraphics[width=\linewidth, center]{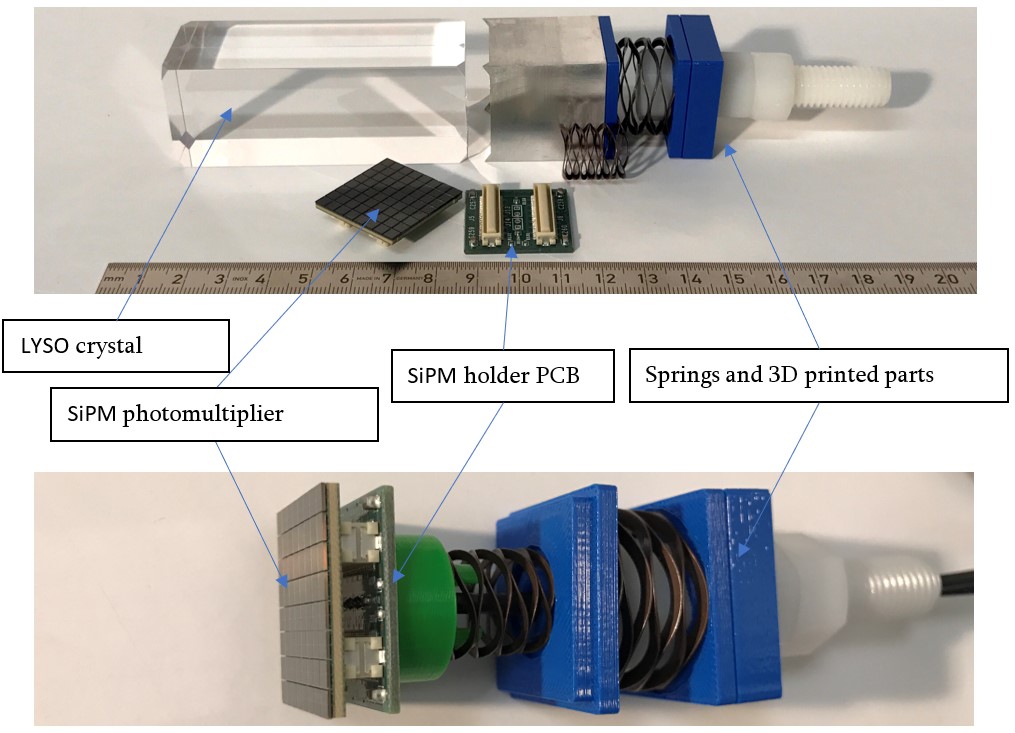}
  \caption{A single calorimeter module before assembly.}
  \label{fig:one module}
\end{figure}

The JEDI (Juelich Electric Dipole moment Investigation) collaboration \cite{www:jedi} develops key-equipment and techniques for the COSY storage ring \cite{art:cosy}, in order to be able to measure the electric dipole moments of protons and deuterons at unprecedented sensitivity \cite{www:YellowReport}. One of the important components of these experiments is the new modular JEDI polarimeter (JePo) \cite{art:keshelashvili, art:keshelashvili_1, art:keshelashvili_2, art:Muller},
consisting of crystal-based calorimeter modules, designed to measure the energy of the particles scattered by the target in the polarimetry reaction. Each module consists of a LYSO (lutetium-yttrium oxyorthosilicate) crystal, a SensL-s J series SiPM (silicon photo-multiplier) \cite{www:SensL}, a SiPM holder PCB, an aluminum holder, a specially designed spring and some other 3D-printed parts, as shown in Figure~\ref{fig:one module}.

In the calorimeter module the scintillation light induced by a particle in the crystal is read out by a SiPM. These types of photon sensors are manufactured in arrays of different size, where each element represents thousands of avalanche photo-diodes (APD), operating in Geiger mode, connected in parallel (see figure 2 in \cite{www:SensL}).

SiPMs are relatively new compared to traditional types of light sensors (e.g. PMT, APD). The process of further development and improvement is continuously being carried out by large manufacturers (Philips, Hamamatsu, SensL, Ketek). Due to the unique features of the SiPMs, such as high gain, high time resolution and their ability to work in strong magnetic fields, these light-sensitive detectors are successfully replacing classic photo-multiplier tubes. The main application areas of SiPMs are PET (Positron Emission Tomography) and hybrid MRI-PET systems, where SiPM-based detectors detect 511 keV $\gamma$-quanta.  In physics experiments at high energies, SiPMs are limited by their finite dynamic range, which is determined by the number of pixels. The energy interval and the number of detected photons in the JEDI experiment are almost three orders of magnitude higher than in the case of PET applications, leading to much higher currents through the SiPMs during the particle detection, while the steady (dark) current remains the same. No multi-channel power supplies are readily available on the market that can provide low output noise level, fast dynamic load, high short-term and long-term voltage stability at the same time, which is needed for an effective operation of the JEDI polarimetry.
The main goal of the R$\&$D work described in this paper was the development of such a multi-channel power supply for SiPM operation of the JePo modules.

\section{Major requirements}
\begin{figure}[ht]
  \centering
  \subfloat[Schematic drawing of SiPM test setup]{\includegraphics[width=\linewidth]{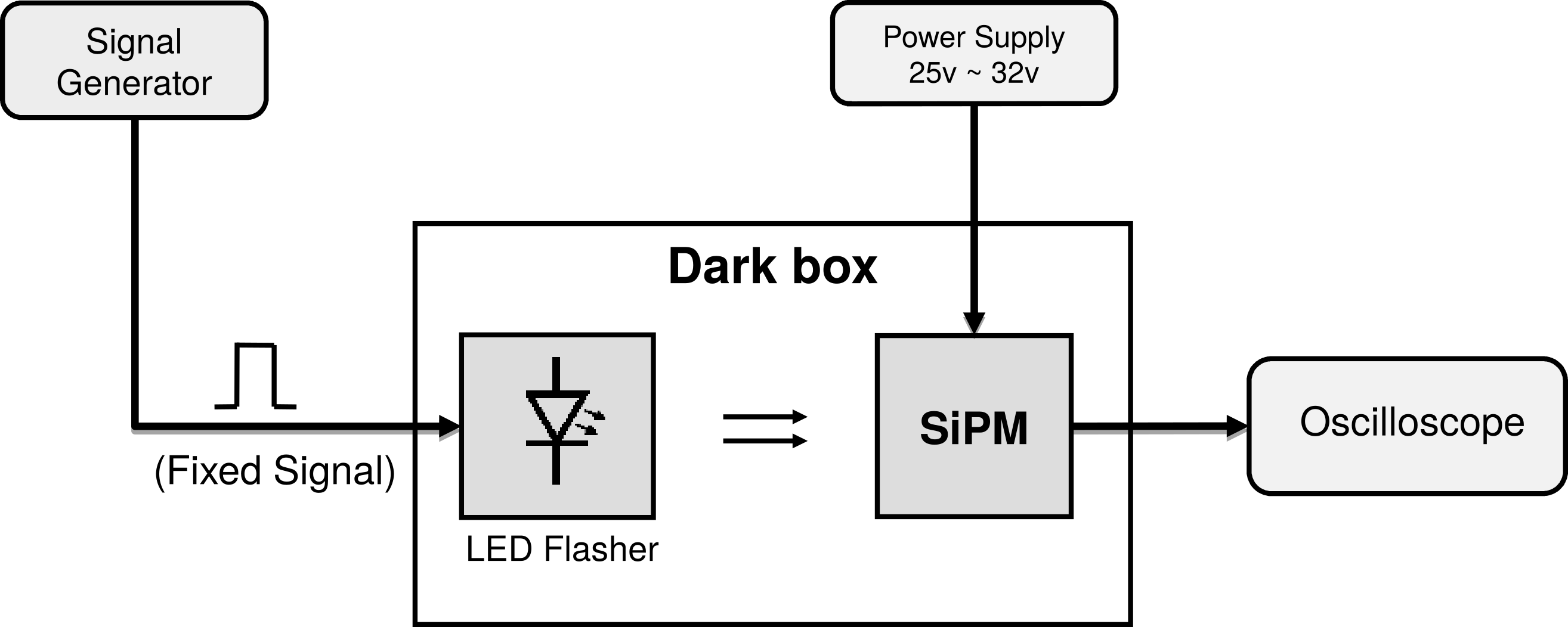}}
  \qquad
  \subfloat[SiPM - Gain as a function of applied voltage]{\includegraphics[width=\linewidth]{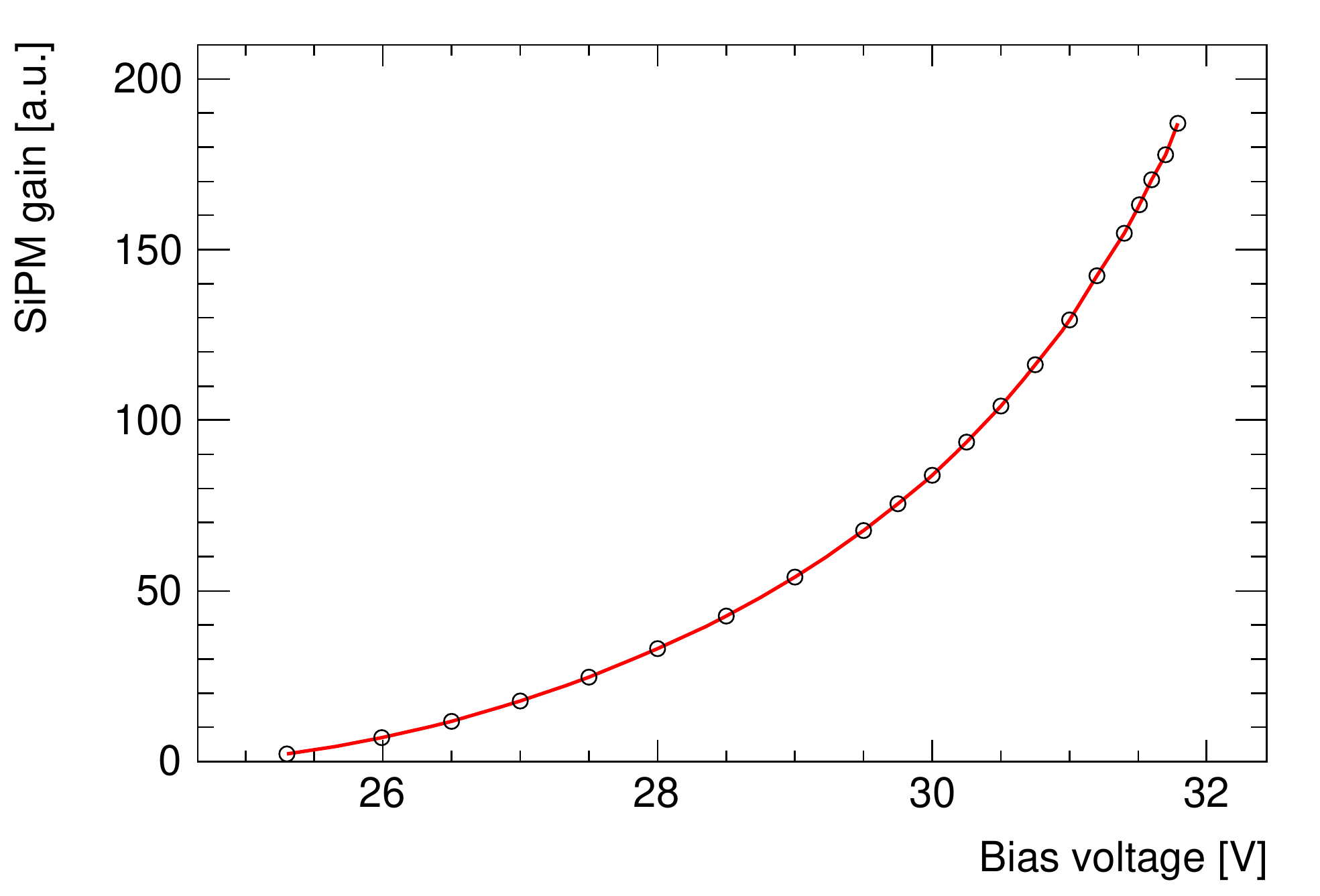}}
  \caption{Test setup for SiPMs (a) and obtained result (b).}
  \label{fig:SiPM_test}
\end{figure}

SiPMs operate at voltages exceeding their breakdown voltage, i.e. in the Geiger mode. In this regime the SiPM gain is, to first approximation, directly proportional to the overvoltage value (difference of the applied reverse voltage and the breakdown voltage). However, we tested our SiPMs in the laboratory to study this behaviour and also check their characteristics at high overvoltage values. For this purpose we put a SiPM array in front of an LED flasher, connected to a signal generator, into a black box (shielding against external light). The output of the SiPM was connected to the digital oscilloscope, as shown in Figure~\ref{fig:SiPM_test} (a), which integrated the electric signal from the SiPM to yield the total electric charge. A signal generator produced rectangular pulses with fixed parameters. By changing the reverse voltage on the SiPM array, we obtained the gain factor as a function of supply voltage, as shown in Figure~\ref{fig:SiPM_test} (b).
As it becomes clear from Figure~\ref{fig:SiPM_test} the gain factor of the SiPM arrays strongly depends on the biasing reverse voltage. Higher gains will lead to an increased SiPM resolution, corresponding to a higher energy resolution of the calorimeter modules. However, the dark current (internal noise) of the SiPM increases faster with voltage. Besides, the energy resolution of the LYSO crystals used in the calorimeter is also limited to about 0.5-1~\% and there is no more benefit from the increase of the SiPM gain too much. As an optimal working voltage range we chose 27 V up to 31 V. Here 1~mV voltage change in the supply voltage induces about 0.05~\% change in the SiPM gain. Therefore, the overall instability plus noise of the power supply output voltage must be less than 10~mV, in order not to allow the overall energy resolution of the calorimeter to degrade due to voltage variations. Hence, it is of high importance to keep the SiPM voltage at a constant level all the time with the best possible short-term and long-term stability and also with low noise.

Extreme variations of the SiPM load present another challenge for the load regulation of the voltage supply due to extreme variations in the current consumption. In the steady state our SiPM array consumes a current of a few $\mu$A, whereas in case of 300~MeV particle detection the peak current through the array reaches 100~mA, which lasts for few micro seconds. If another particle hits the same calorimeter module within the voltage restoration time period, which is particularly the case for high counting rates, it will be misinterpreted as a less energetic particle, finally also resulting in a degraded energy resolution of the calorimeter.

The gain factors of different calorimeter modules are to match each other. This requires the voltage for each supply module to be adjustable. Therefore, the power supply needs to be modular with as many channels, as there are separate detector modules in the polarimeter detector (around 200). In addition, this approach also overcomes the potential cross talk problems between the detector modules via the supply lines.

Further requirements include remote on/off switching, voltage adjustment, current and temperature monitoring features.

Considering the above-mentioned requirements, we developed the new power supply modules based on a linear voltage regulator.  Apart from its simplicity and reliability the output voltage is free from high-frequency spikes characteristic for DC-DC converters.

\section{Power Supply Modules}
As already mentioned, by analysing the main requirements for the power supply, the supply modules were decided to be linearly regulated. Initially, few readily available voltage regulator ICs were considered for that purpose and their features and capabilities had been analysed in detail. Although not the easiest one to use, a precision voltage regulator IC - UA723 \cite{www:Ti_regulator} in DIP-14 package from Texas Instruments has been chosen. While operating at voltages up to 40~Volts it can control the output at up to 37~Volts, delivering currents of up to 150~mA to the load.

\begin{figure}[ht]
  \includegraphics[width=\linewidth,center]{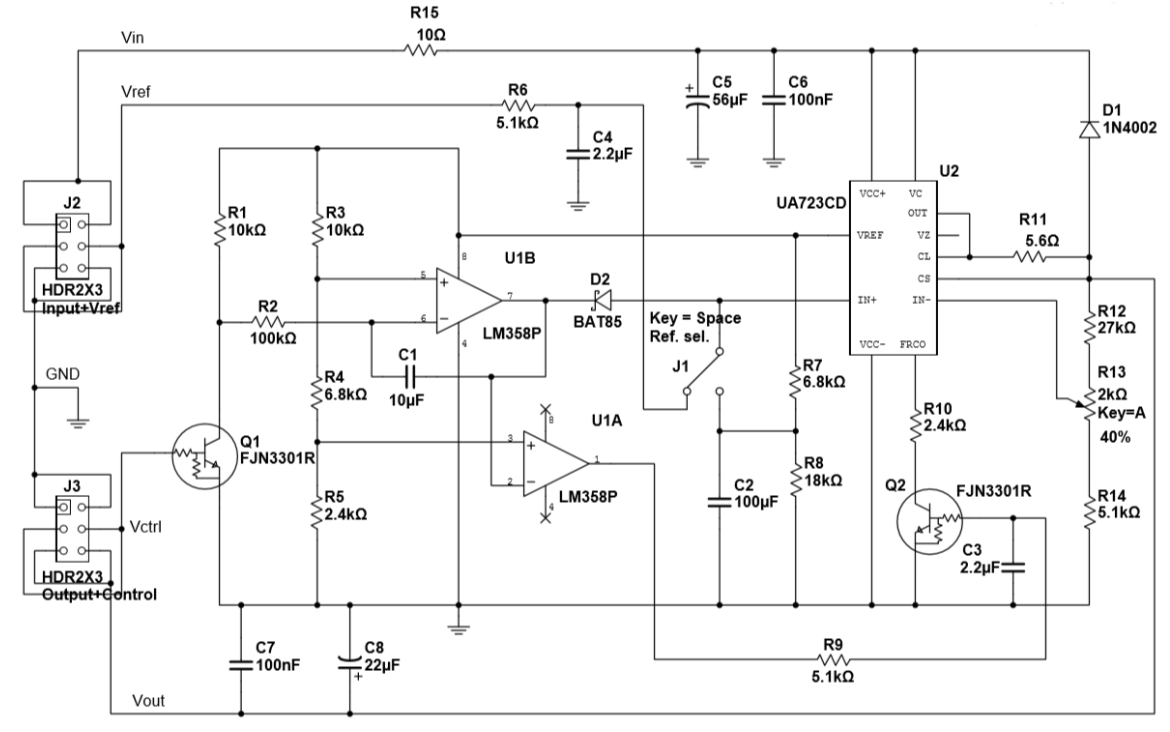}
  \caption{The schematics of a single module of the power supply}
  \label{fig:powersupply}
\end{figure}

The schematics of the power supply module is shown in Figure~\ref{fig:powersupply}.
Two additional features were added to the basic UA723 external circuit; a remote on/off switching and a slow voltage ramp up/down during switching. These features are realised using a simple dual operational amplifier \cite{www:Ti_dualAmp} and two NPN transistors with internal bias to reduce the number of external components. One of the two operational amplifiers (U1B in the schematics) operates as an integrator, providing linear ramp up and ramp down of its output. Being decoupled from the main regulator IC by a diode, it pulls down the 5~V reference voltage input to the ground potential, dragging the output of the regulator voltage proportionally also towards zero. However, due to the internal limitation of the UA723 IC, the output voltage differs from zero even if the reference input is grounded. For that reason an additional circuit with another operational amplifier (U1A) and transistor is added. Performing like a comparator with an open collector, it blocks the regulator output when the ramp integrator output voltage falls below a specific voltage (close to zero). During the ramp-up phase, since the operational amplifier is powered by about 7~Volts, the output of the integrator goes above the reference voltage level and decouples from it by means of a diode D2. In such a way, since the reverse current of the chosen diode is negligible, the ramp integrator stage of the circuit has no influence on the regulator during an active operation of the supply module.

The output voltage can be precisely adjusted via a multi-turn potentiometer R13. The low temperature coefficient resistors R12 and R14 are selected in such a way that the voltage can be adjusted roughly from 26~V up to 33~V. The maximum load current is limited to approximately 120~mA by a resistor R11. The activation of the power supply is performed by applying +3.3~V to the base of the Q1 transistor, which is delivered by the Raspberry Pi computer, remotely controlling the whole power supply and its voltage monitoring system, as will be reported later in this article.

\subsection{Voltage reference}
A major problem with the selected regulator IC found during the laboratory test is related to the internal voltage reference. The output noise voltage level reported in the product data sheet in a frequency range of 100~Hz to 10~kHz is rather small. However, long-run tests revealed much poorer stability at longer time periods, affecting the short-term and long-term stability of the power supply. Fortunately, the chosen IC can also be operated by an external reference voltage. For this purpose, we constructed a single reference board, based on a very stable 5~V voltage reference IC - MAX6350 \cite{www:MAX63}, to serve all the power supply modules. The selection of the internal/external reference is possible via a J1 jumper (see schematics in Figure~\ref{fig:powersupply}).

\begin{figure}[ht]
  \includegraphics[width=\linewidth, center]{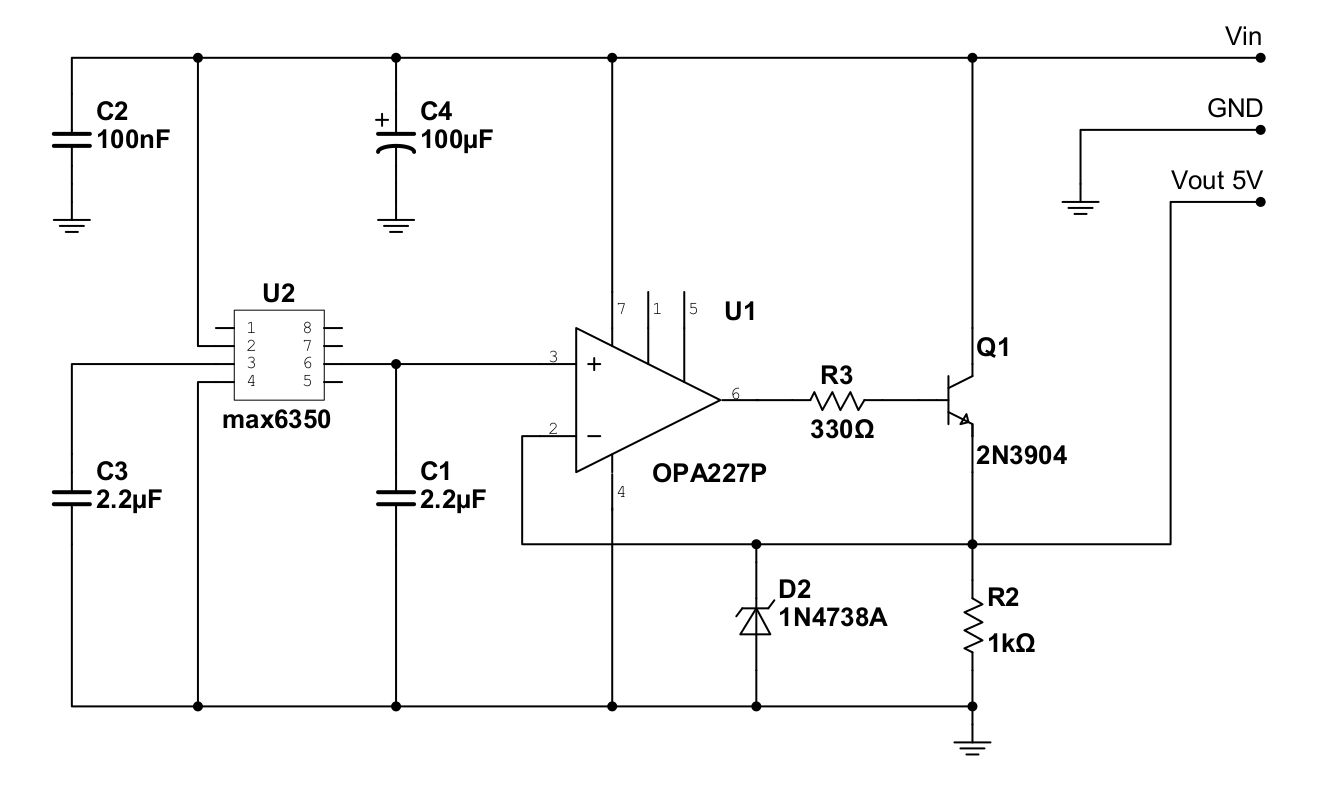}
  \vspace{-15pt}
  \caption{The schematics of a reference voltage source}
  \label{fig:reference}
\end{figure}

Each power supply module consumes around 1~mA of current from the reference input pin when turned off. The final detector will include more than 100 such modules. To handle the total current an additional current buffer, consisting of an operational amplifier and an output transistor, has been added to the output of the reference IC, as shown in the schematics in Figure~\ref{fig:reference}. For this purpose the operational amplifier OPA227 \cite{www:Ti_prec_opamp} from Texas Instruments was chosen, featuring low-noise and very small offset voltage drift with temperature. Although the operational amplifier together with the output transistor provides a unit voltage gain, its high open-loop gain of 160~dB ensures a good load regulation, which is very important since the reference is shared among all the supply modules.
The power supply rejection ratio of this device is also high, which is beneficial for maximally decoupling the output from the main supply.

\subsection{SiPM load response}
\begin{figure}[ht]
\centering
  \includegraphics[width=0.7\linewidth]{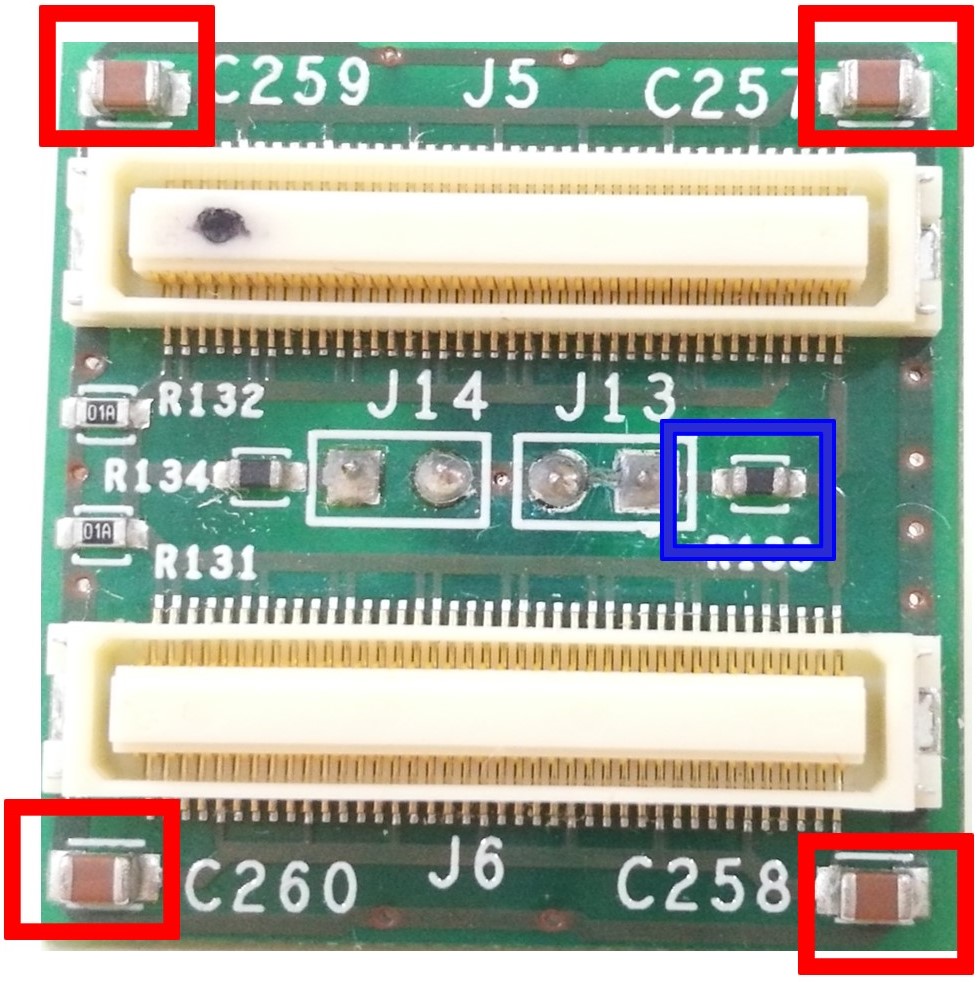}
  \caption{SiPM original board with LC filter. Red squares indicate the buffer capacitors while the blue square shows the filtering inductance, which was present in the original design of the board and was removed in later versions.}
  \label{fig:holder_board}
\end{figure}

The calorimeter modules of the polarimeter detector are connected to the power supply via ca. 10 meter long LEMO cables. In this case the voltage drop in the cables during the particle detection becomes significant. To overcome this problem, multi-layer ceramic capacitors for energy buffering are added to the SiPM supporting boards, as shown in Figure~\ref{fig:holder_board} with red squares. They provide electric charge during the signal rise time and are recharged again by the power supply through the LEMO cable after the signal. These supporting boards have been designed and assembled at Ferrara University (Italy) \cite{www:Ferrara}. They originally contained four capacitors,  2.2~$\mu$F each. 

In order to test whether the power supply modules together with SiPM boards performed well, a special apparatus had been developed to simulate SiPM loads under laboratory conditions without particle beam and without calorimeter modules.

\begin{figure}[ht]
  \subfloat[]{\includegraphics[width=\linewidth]{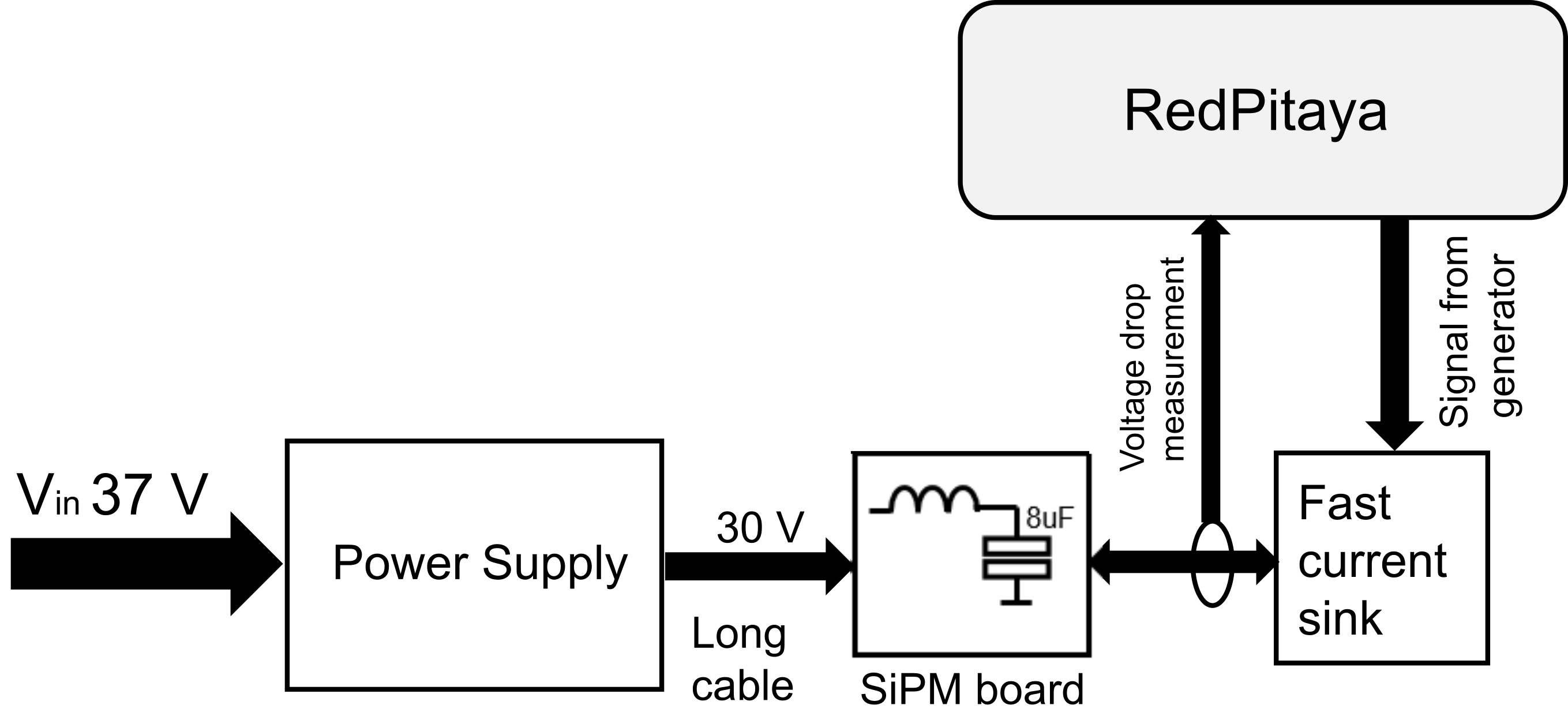}}
  \qquad
  \subfloat[]{\includegraphics[width=\linewidth]{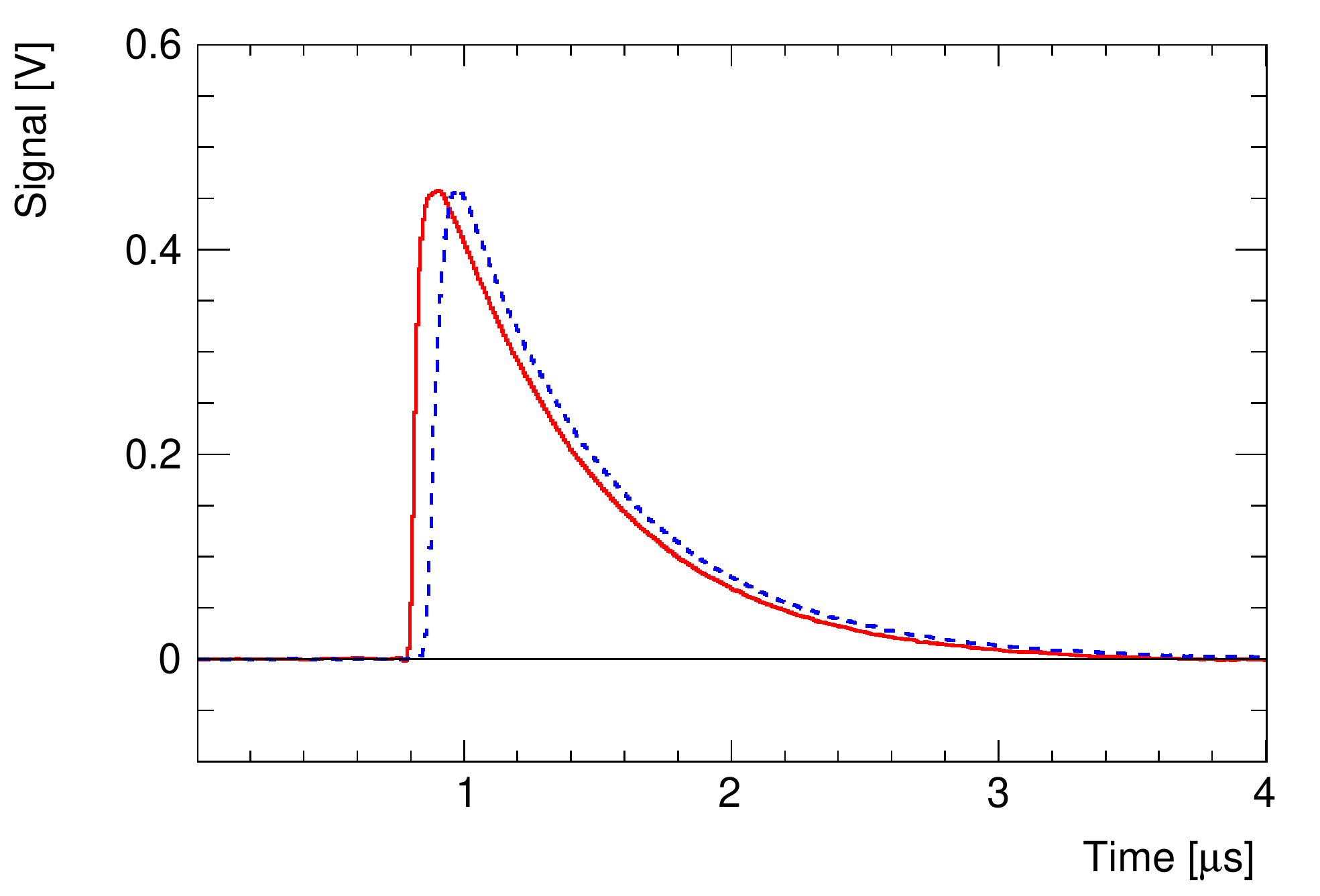}}
  \caption{ (a) Laboratory setup for a SiPM load simulation test and (b) comparison of the shapes of the Red Pitaya-generated voltage signal (solid red) versus the simulated sink current, converted back to voltage (dashed blue). The simulator consists of Red Pitaya signal generator, providing the exact copy of the SiPM signal and the fast voltage to current transformer loading the output of the power supply module. The voltage drop is measured at the output of the SiPM holder board and is fed back into the Red Pitaya oscilloscope.}
  \label{fig:Dv_measurement}
\end{figure}

As shown in Figure~\ref{fig:Dv_measurement} (a) the test setup consists of the power supply module connected to the supporting board via a LEMO cable. Instead of the SiPM there is a specially designed voltage-regulated fast current sink, which sinks the current from the supply proportionally to the input voltage. The voltage signal to the current sink is delivered via the Red Pitaya \cite{www:RedPitaya} board, which can be used as an arbitrary function generator and an oscilloscope simultaneously, thanks to its dual input and output channels.

The Red Pitaya signal generator was programmed to precisely reproduce the voltage signal from the calorimeter module. A standard signal generator software available on the Red Pitaya system was modified to accept arbitrary signal shapes representing the experimental data recorded during previous test runs.

As already noted a fast current sink was designed to convert the simulated SiPM voltage signal to the load current. In order to tune the different parameters of the converter, feedback loops and filters, simulation software such as "Altium designer" \cite{www:altium} and "NI Multisim" \cite{www:multisim} has been used. After few iterations an acceptable result was obtained. In Figure~\ref{fig:Dv_measurement} (b) an original voltage signal is shown together with the simulated current (converted back to voltage) signal. Apart from a delay of about 70~ns due to the converter electronics and the connecting cables, the amplitudes and the shapes of the original (solid red) and converted (dashed blue) signals are almost identical. The delay does not affect any of the test procedures described further below, since the absolute times of the simulated pulses play no role at all.

The input channels of the Red Pitaya board were used to monitor the AC content of the SiPM board output (i.e. the SiPM supply voltage). Although the standard built-in oscilloscope web application of the Red Pitaya could also have been used, we applied previously developed data acquisition software based on the CERN ROOT libraries instead \cite{www:CERN}. This permitted us to take a closer look at the obtained results in the offline regime later.

\begin{figure}[ht]
  \includegraphics[width=\linewidth, center]{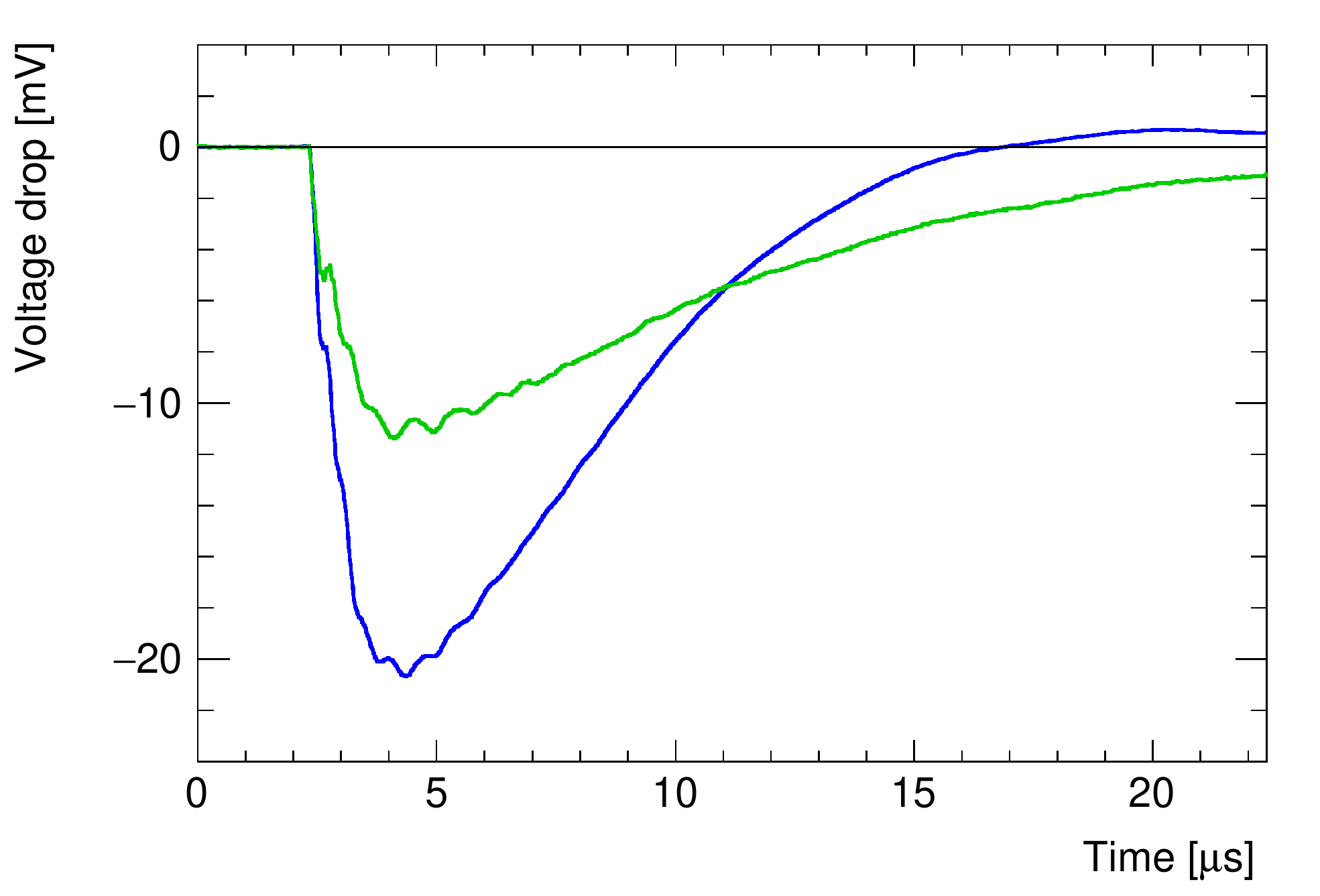}
  \caption{Voltage drop at the SiPM board (SiPM side): blue curve - voltage drop on the 
  first prototype of the SiPM board with 8~$\mu$F total capacitance, green curve - voltage drop 
  on an improved board with the capacitance increased up to 40~$\mu$F.}
  \label{fig:Dv_graph_joined}
\end{figure}

The SiPM load simulation was performed for different scenarios corresponding to different experimental conditions: high and low count rates, bursts, particles with higher and lower energies, pile-ups and etc. During the test the voltage drops were measured at both ends of a long cable. However, most of all we were interested in voltage variations at the SiPM side. During the short load pulses most of the energy is drawn from the capacitors on the SiPM board. Therefore, in order to ensure a stable supply voltage for the SiPMs, a minimum capacitance should be provided on the boards. In the case of four 2.2~$\mu$F capacitors a voltage drop of 21~mV was measured, as shown in Figure~\ref{fig:Dv_graph_joined} (blue). After increasing the capacity and reducing the internal resistance and inductance of the board, we measured a new version of the board with a total capacitance of up to 40~$\mu$F. This reduced the voltage drop down to about 12~mV (in red).

\section{Voltage monitoring and controlling system}
The stability of the supply voltage plays an important role in the overall performance of the calorimeter. In order to control the behaviour of the power supply and track its output, a separated multi-channel voltage monitoring system has been developed and built.

Since we are dealing with voltages around 30~V and an accuracy around $0.1$~mV, initially, multi-channel ADCs with high input voltages and high resolution had been considered. However, the selection of such converters on the market is very limited. Therefore, we decided to build a multiplexed $i.e.$ sequential readout system with a single output channel. The developed 128-channel system is a separate device, which is connected to supply modules and can select channels sequentially or selectively. The voltage is then measured by the high-precision desktop multimeter "Keithley 2700" \cite{www:Keithley}, providing a resolution of 100~$\mu$V. The multimeter is controlled by a Raspberry Pi micro computer via a UART interface. 

\begin{figure}[ht]
  \subfloat[]{\includegraphics[width=\linewidth]{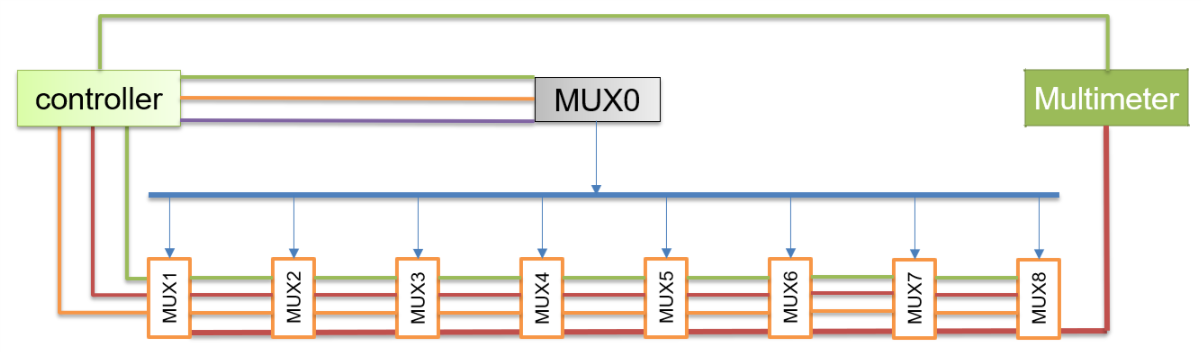}}
  \qquad
  \subfloat[] {\includegraphics[width=\linewidth]{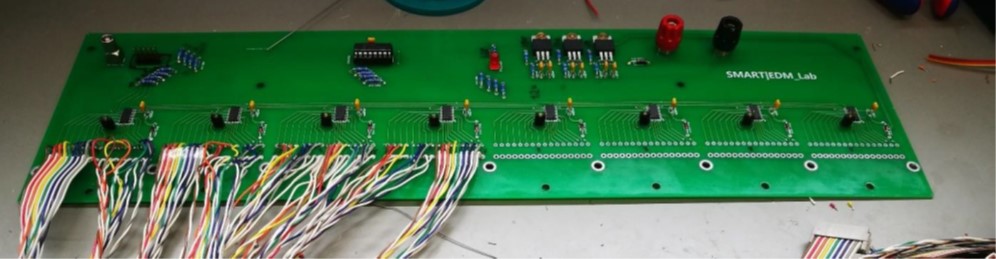}}
  \caption{(a) A block diagram and (b) an assembled board of the voltage monitoring system}
  \label{fig:Monitoring}
\end{figure}

The monitoring system board consists of eight 16-channel high-voltage multiplexers, as shown in Figure~\ref{fig:Monitoring} (a). MUX0 is an 8-channel multiplexer that has 1 activation pin and 3 control pins, and MUX 1-8 a 16-channel multiplexer with 1 activation and 4 control pins. The controller allows MUX0 to activate only one of the eight MUX1-MUX8 at a time by enabling or disabling their activation pins. The monitoring system board is powered by three linear voltage regulators LM317HV \cite{www:LM317}. The first one regulates the input voltage of 40~V down to 32~V supplied to the 16-channel multiplexers. The second regulator further reduces the voltage down to 16~V and the third one - from 16~V to 5~V, which is used as a supply for the 8-channel multiplexer. The board has an LED status indicator connected to the third regulator, and it works only if all the regulators work properly.

% \begin{figure}[ht]
%   \includegraphics[width=\linewidth, center]{pics/17.png}
%   \caption{A block diagram of the voltage monitoring system}
%   \label{fig:ref_diagram}
% \end{figure}

The parameters of many multiplexers have been studied in detail to achieve a high accuracy and a high stability of the measuring system. Two types of parameters can be identified. Parameters of the first type affect all the channels of the system in a systematic way, whereas those of the second type introduce variations between the channels. Documented maximum values of the parameters were used to estimate their effects on the voltage readout accuracy. The {\it on-state} resistance of the multiplexed channels averages to $R_{on}\approx140~\ohm$. The smaller the value, the more accurate is the voltage at the output. This parameter introduces a constant reduction of the output, which is approximately 5~mV. A more important parameter is the maximum deviation of the impedance from the mean value - $\Delta R_{on} \approx 4~\ohm$. Its maximum impact on the output voltage is about 0.3~mV. Another important parameter of the multiplexer is the average leakage current of an {\it off-state}, in the order of $\pm$ 100~pA. There are eight multiplexers in the system. Therefore, the maximal effect of the total leakage current on the 1~M$\ohm$ output resistor is less than 1~mV. Another parameter is the maximum deviation of the leakage current from the mean value - $\Delta$Id$\approx$10~pA, with a maximal impact on the output voltage of less than 0.1~mV. Depending on the given parameters, the difference between the channels will be less than 0.5~mV. The cross talk between each group of 16 channels (within one multiplexer IC) is -75~dB at 1~MHz, which can become important in case if a high-speed device, for instance an oscilloscope, is connected to the output to measure the high-frequency content in the detector supply lines. 

Although the values of the above-mentioned parameters are available in \cite{www:AD_mux}, a set of laboratory measurements has been performed in order to ensure the correctness of the documented data, and to also measure the parameters in operating regimes, where the information was lacking. Finally, we identified the ADG5206 \cite{www:AD_mux} multiplexers from Analog Devices as a good candidates for this task. There is also an additional 8-channel multiplexer - DG408 \cite{www:VS_mux} from Vishay Siliconix, which enables only one out of the eight readout multiplexer ICs at a time. An assembled monitoring system board before the installation is shown in Figure~\ref{fig:Monitoring} (b).

% \begin{figure}[ht]
%   \includegraphics[width=\linewidth, center]{pics/19.jpg}
%   \caption{An assembled board of the voltage monitoring system}
%   \label{fig:ref_pic}
% \end{figure}

As mentioned above, more than 100 calorimeter modules are supposed to be used in the experiment and each has its own power source with turn on/off function. These modules as well as the voltage monitoring system must be controlled by a remote computer. A dedicated Raspberry Pi mini computer was selected for this task. A special software, based on Python, has been created, which provides online control of the modules, monitors the voltages and records the voltage data for further analysis. It consists of several parts: the voltage monitoring, the power supply channel control and the web server. The voltage monitoring part of the software reads 128 channels in sequence or any particular selected channel. It also controls the output multimeter in parallel and sends the read data to the server once the scan of all the channels is completed. The server part is also written in Python and runs on the same Raspberry Pi. It provides the graphical user interface via a web browser and also allows control from any remote computer within the same network.

For controlling the power supply modules, the available Raspberry Pi I/O pins are not enough to control 100+ channels. To solve this problem, 16-channel bidirectional port extenders - MCP23S18 \cite{www:port_expander} from Microchip, working on SPI interface, are used. 

\section{Power supply assembly}
The whole multi-channel power supply consists of the following parts: voltage regulator modules, reference voltage source, voltage monitoring system, port expander and Raspberry Pi. There are additional breakout boards - motherboards to connect different systems. Each motherboard accommodates eight voltage modules and has connectors for on/off control, reference voltage and the monitoring system (connectors J2 and J3 on the schematics in Figure~\ref{fig:powersupply}). The output conectors on the front panel of the power supply are distributed in the shape of the real detector for easy cabling and maintenance. The power supply control software was developed with the same approach, also representing the front view of an assembled detector, and providing live voltage control and monitoring features.

\begin{figure}[ht!]
  \includegraphics[width=\linewidth, center]{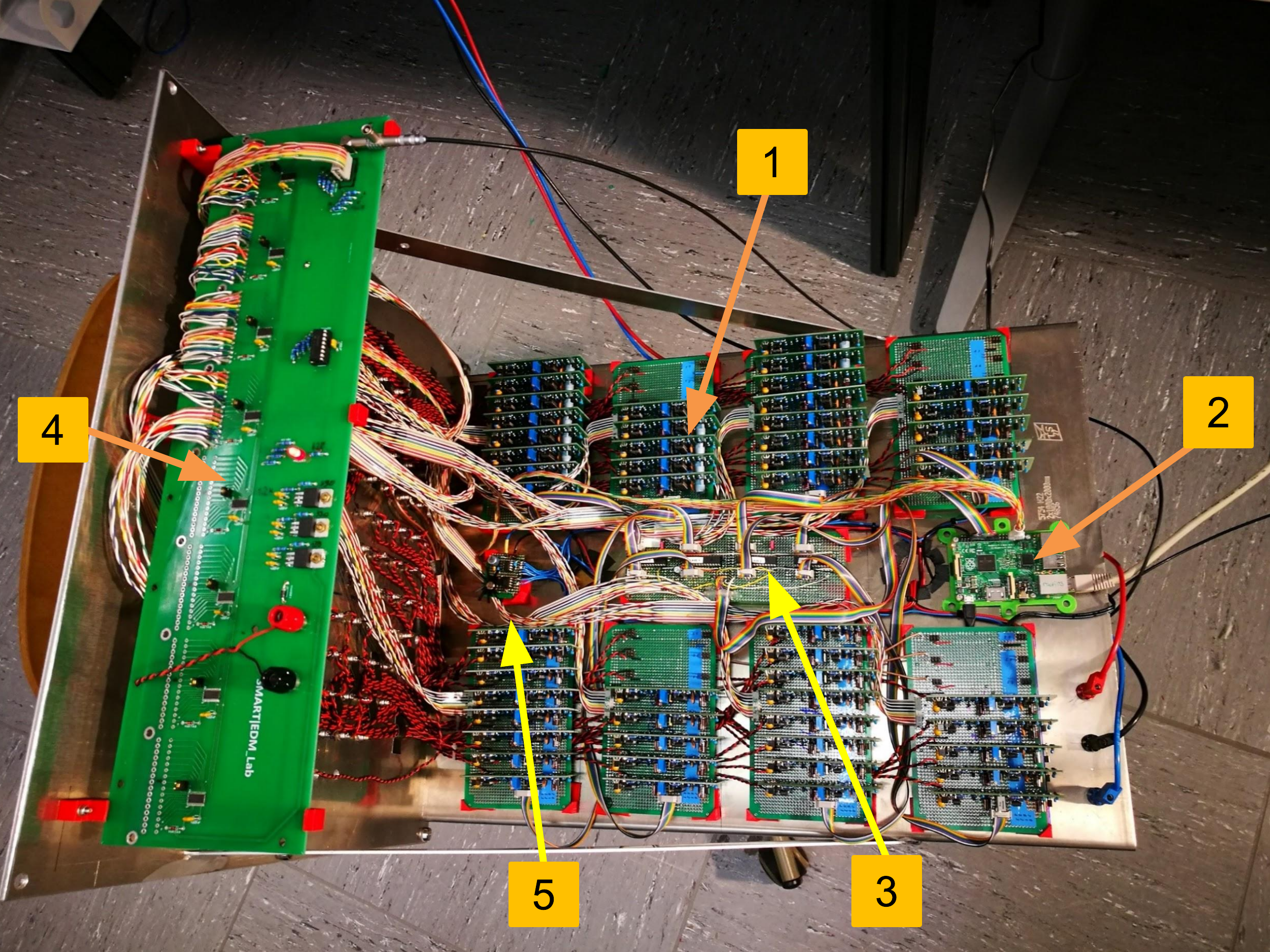}
  \caption{Assembled power supply: 1 - Power supply modules,
  2 - Raspberry Pi, 3 - Port expander, 4 - Voltage monitoring board, 
  5 - Reference voltage board}
  \label{fig:real_PWS}
\end{figure}

The fully assembled power supply is shown in  Figure~\ref{fig:real_PWS}.
\begin{figure}[ht!]
    \subfloat[]{\includegraphics[width=\linewidth]{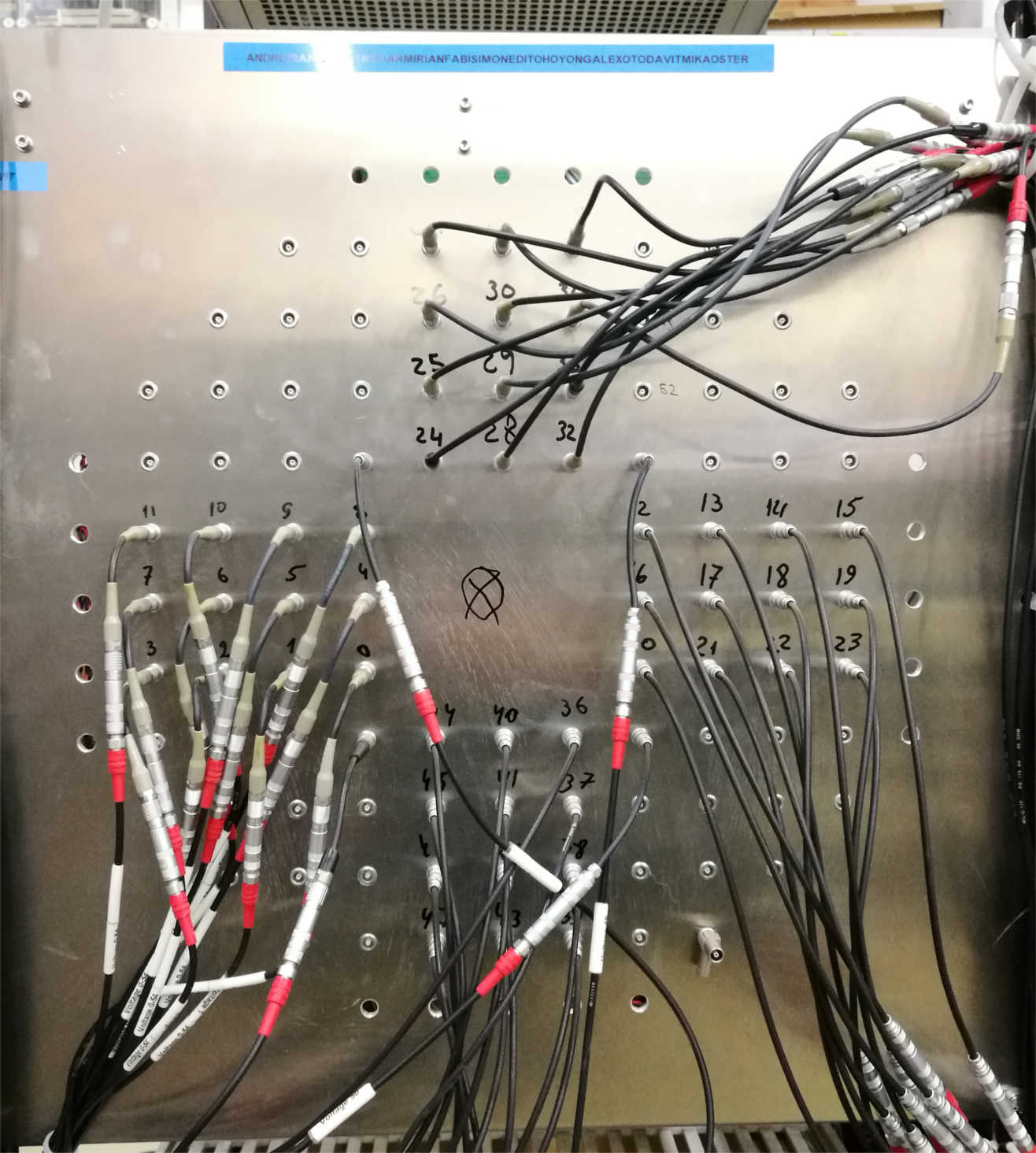} }
    \qquad
    \subfloat[]{\includegraphics[width=\linewidth]{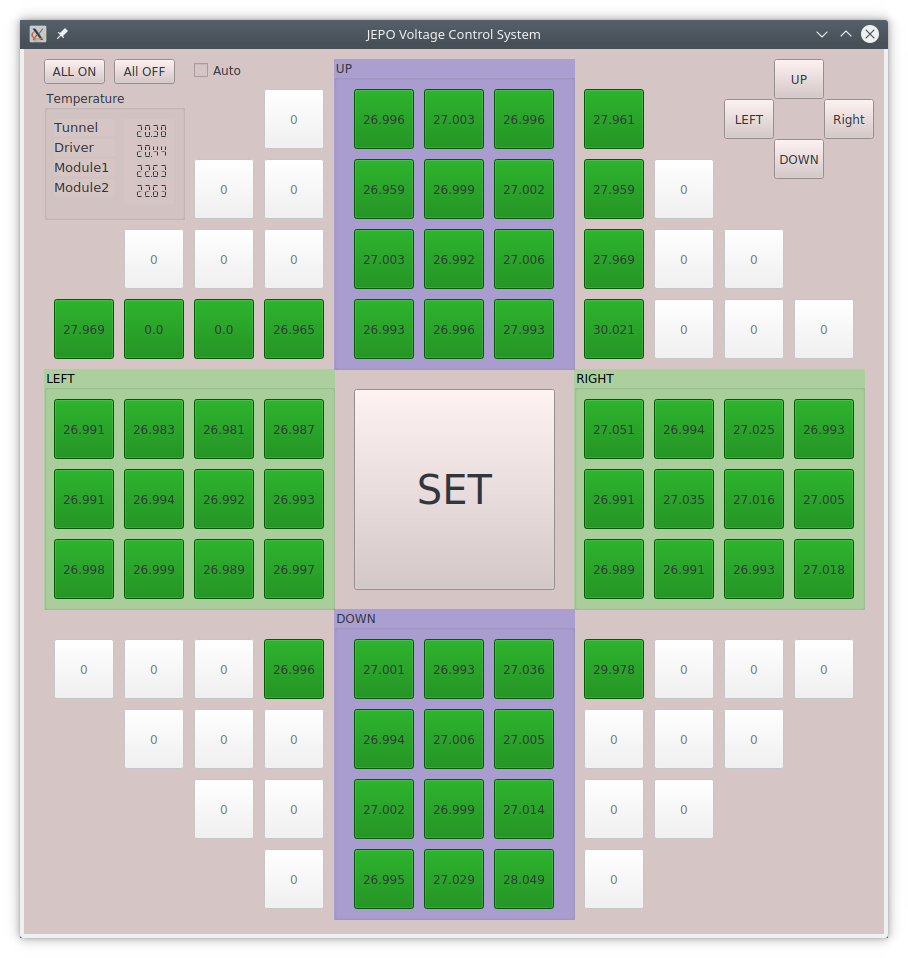} }
    \caption{ (a) A front view and (b) a control panel of the slow control software for the power supply}
    \label{fig:ps_view}
\end{figure}

\section{Test runs and performance analysis}
The system was carefully examined during the test-experiments of the polarimeter, and data have been recorded to analyse its performance and further improve some of the characteristics. In general, the output voltage may vary due to temperature changes, internal noise and also due to variable loads. Variable load conditions, which are associated with varying intensities of detected particles from cycle to cycle, have been studied by using voltage monitoring data in conjunction with the calorimeter data obtained during one of the test experiments on an external COSY beam. As a result, no substantial correlation between the voltage variation and the counting rates has been found. More details on this topic can be found in the Master's thesis by O.~Javakhishvili~\cite{thes:oj}. As was identified later, voltage stability was mainly affected by the change in environmental conditions like the temperature.

\comment{
\begin{figure}[ht]
  \includegraphics[width=\linewidth, center]{22.jpg}
  \caption{Original voltage data of a single channels}
  \label{fig:sig_original}
  \includegraphics[width=\linewidth, center]{23.jpg}
  \caption{Frequency spectrum of the voltage data}
  \label{fig:sig_frequency}
\end{figure}
\begin{figure}[ht]
  \includegraphics[width=\linewidth, center]{24.jpg}
  \caption{Low-pass-Filtered voltage signal of the same channel}
  \label{fig:sig_filtered}
  \includegraphics[width=\linewidth, center]{25.jpg}
  \caption{High frequency content of the voltage data $v(t)$ - $\overline{v(t)}$ }
  \label{fig:sig_noice}
\end{figure}
}

\begin{figure}[ht]
  \includegraphics[width=\linewidth, center]{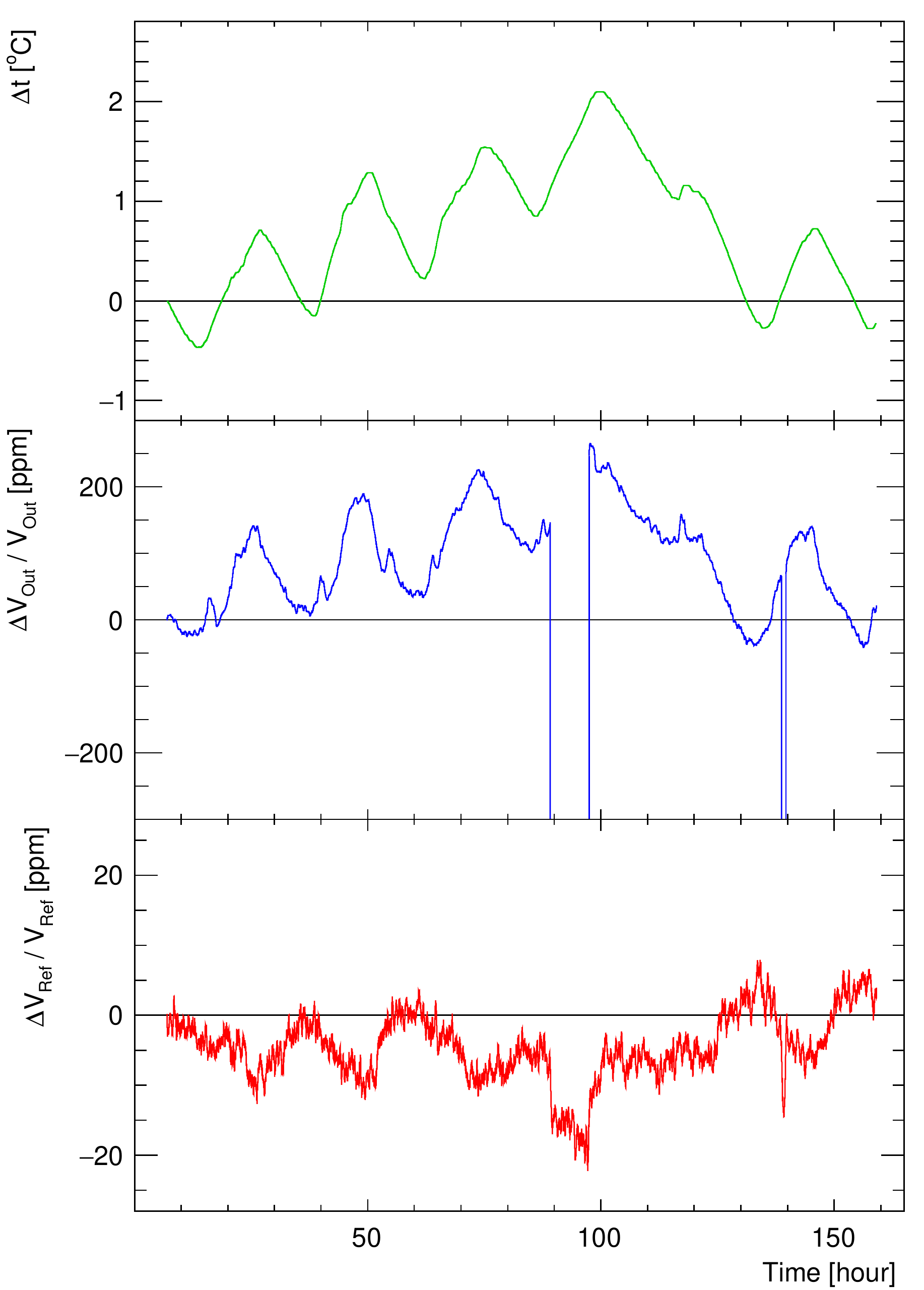}
  \caption{Voltage stability as function of time. The output voltage of one of the power supply modules is shown (middle) together with the reference voltage (bottom) in parallel with the ambient temperature recording (top). Note that the drops in the output voltage are caused by turning the power supply modules off. The reference voltage source remains turned on all the time.}
  \label{fig:supply_stability}
\end{figure}

Longer records of the voltage values revealed a strictly periodic behaviour of the peaks in the voltage data suggesting the effect to be associated with day-night cycles. This was later proved to be true by investigating the ambient temperature using digital thermometers and correlating the temperature and voltage data. This is demonstrated in Figure~\ref{fig:supply_stability}: the behaviour of the output voltage (middle) follows the ambient temperature variation (top), suggesting the significantly positive temperature coefficient. It must be noted, that the reference voltage output (bottom) remains much more stable meanwhile.

\comment{
\begin{figure}[ht]
  \includegraphics[width=\linewidth, center]{27.jpg}
  \caption{8 days voltage data of one channel. black-unfiltered, red-filtered}
  \label{fig:8day_voltage}
  \includegraphics[width=\linewidth, center]{28.jpg}
  \caption{8 days data of reference voltage. black-unfiltered, red-filtered}
  \label{fig:8day_ref}
\end{figure}
}
In order to investigate the temperature dependence of the single module of the power supply in detail, a temperature chamber with the outer dimensions of  $50\times50\times50$~cm$^{3}$ has been built by using the 6~cm thick insulating material XPS, which is mainly used in building construction as a thermal insulator material. The temperature chamber was equipped with a cooler - Peltier-element as well as a heater device, which was assembled using high-power resistors and a passive radiator. This allowed the chamber to sustain temperatures in its internal volume ranging from 15$^{\circ}$C to 35$^{\circ}$C at a room temperature of about 23$^{\circ}$C. The voltage reference source was identified to be only slightly sensitive to the temperature. Its temperature coefficient was estimated to be around 4~$\mu$V/K at 5~Volts (0.8~ppm/K). The power supply module together with the voltage reference source was measured to have a high temperature coefficient of about $\approx$~3.4~mV/K at 28~Volts. In order to minimise the temperature coefficient, an attempt was made to analyse the schematics and to identify electric components, significantly affecting the temperature stability. The effects of these components have been identified using the temperature chamber. Afterwards these components were replaced by much more stable substitutes.

\begin{figure}[ht]
  \includegraphics[width=\linewidth, center]{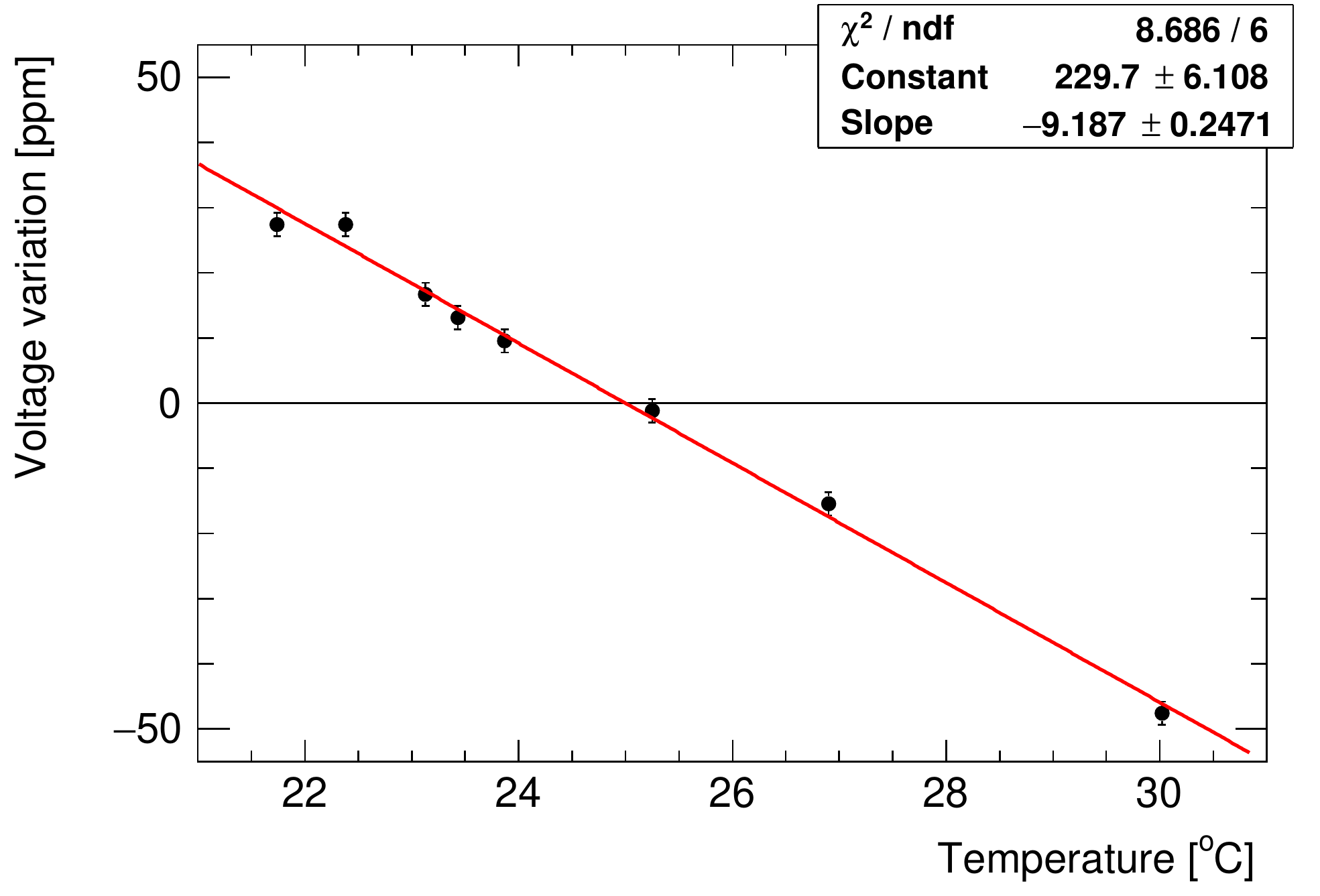}
  \caption{Temperature dependency of the power supply module after the upgrade.}
  \label{fig:temperature}
\end{figure}

\begin{figure}[ht]
  \includegraphics[width=\linewidth, center]{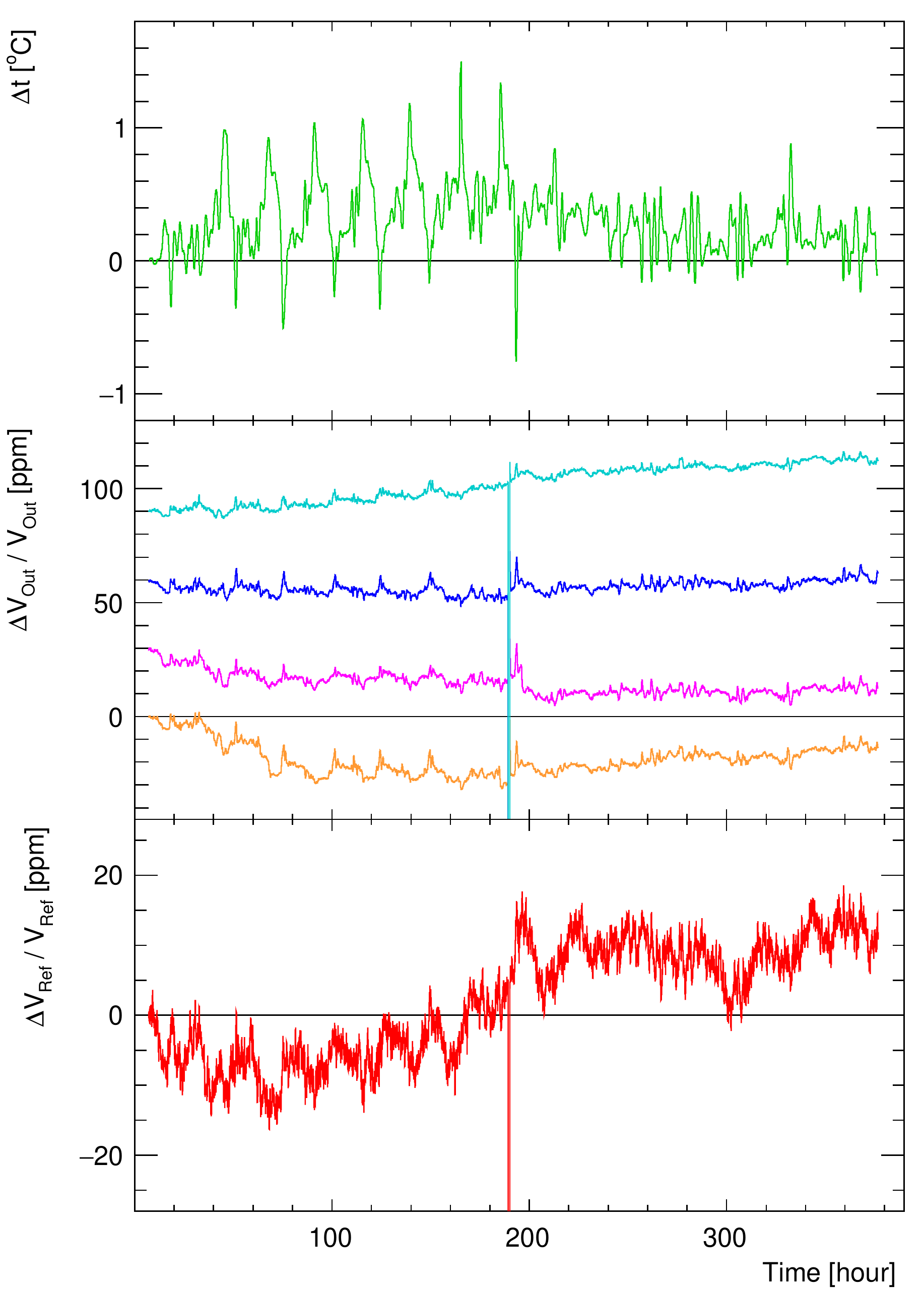}
  \caption{Voltage stability of the upgraded power supply. The output voltages of four typical power supply modules are shown (middle) together with the reference voltage (bottom) and the ambient temperature recording (top). Note that the whole power supply was off during some period of time. This is indicated by a drop in the output voltages at around 190 hours.}
  \label{fig:supply_stability_2}
\end{figure}

After upgrading the power supply modules, a scan was performed to measure the resulting temperature coefficient. As demonstrated in Figure~\ref{fig:temperature} the resulting thermal stability was improved by more than an order of magnitude; the temperature coefficient of one particular module under test was changed from +3.4~mV/K to -0.25~mV/K at 28~Volts (from +120~ppm/K to -9~ppm/K). The above mentioned final numbers turned out to be typical values also for all other modules, as demonstrated in Figure~\ref{fig:supply_stability_2}.

\section*{Summary and outlook}
In this paper, the development of a multi-channel power supply for SiPMs has been described, including a network-based control and monitoring system for each individual channel and the corresponding user-friendly software. 
Tests were performed both in the laboratory and in the accelerator beam environment. As the most striking development result, the temperature stability was improved from about 120~ppm/K to approximately 9~ppm/K. In order to further improve the power supply, several new features are planned to be added, including remote voltage adjustment, temperature and current consumption monitoring for each channel. This will allow us to modify operation conditions online. Moreover, the high stability of the power supply modules provide us with the possibility to easily implement the slow software feedback control if necessary. For instance, such a technique can be very useful to correct for the temperature-dependent calorimeter parameters, such as SiPM gain and the light output of a LYSO crystal.

\section*{Acknowledgements}
The authors wish to thank the members of the JEDI collaboration for their support concerning the technical aspects of using the current device in the experiment at COSY. This work has been financially supported by the research center J\"ulich via COSY-FFE program, by an ERC Advanced~-~Grant (srEDM \#694340: "\textit{Electric Dipole Moments using storage rings}") of the European Union, and by the Shota Rustaveli National Science Foundation of the Republic of Georgia (SRNSFG) grant No. 04/01: "\textit{Search for Electric Dipole Moments using Storage Rings (srEDM)}".

\bibliography{bib}

\begin{thebibliography}{10}
\expandafter\ifx\csname url\endcsname\relax
  \def\url#1{\texttt{#1}}\fi
\expandafter\ifx\csname urlprefix\endcsname\relax\def\urlprefix{URL }\fi
\expandafter\ifx\csname href\endcsname\relax
  \def\href#1#2{#2} \def\path#1{#1}\fi

\bibitem{art:sakharov}
{A.~Sakharov}, Violation of $cp$ invariance, $c$ asymmetry, and baryon
  asymmetry of the universe, JETP Letters 5 (1967) 24.

\bibitem{www:jedi}
{JEDI Collaboration}, {Homepage},
  \url{http://collaborations.fz-juelich.de/ikp/jedi/} (2019).

\bibitem{art:cosy}
{R.~Maier et al.}, Nuclear Instruments and Methods A 390  1, (1997).

\bibitem{www:YellowReport}
{F.~Abusaif et al.}, {Storage Ring to Search for Electric Dipole Moments of
  Charged Particles, CERN-PBC-REPORT-2019-002},
  \url{https://arxiv.org/pdf/1912.07881.pdf} (2020).

\bibitem{art:keshelashvili}
I.~Keshelashvili, F.~Müller, D.~Mchedlishvili, D.~Shergelashvili, {A new
  approach: LYSO based polarimetry for the EDM measurements}, J. Phys. Conf.
  Ser. 1162~(1) (2019) 012029.
\newblock \href {https://doi.org/10.1088/1742-6596/1162/1/012029}
  {\path{doi:10.1088/1742-6596/1162/1/012029}}.

\bibitem{art:keshelashvili_1}
I.~Keshelashvili, F.~Müller, D.~Mchedlishvili, D.~Shergelashvili, {Polarimetry
  - from basics to precision}, PoS PSTP2017 (2018) 025.

\bibitem{art:keshelashvili_2}
I.~Keshelashvili, F.~Müller, D.~Mchedlishvili, {Polarimetry concept based on
  heavy crystal hadron calorimeter}, J. Phys. Conf. Ser. 928~(1) (2017) 012018.
\newblock \href {https://doi.org/10.1088/1742-6596/928/1/012018}
  {\path{doi:10.1088/1742-6596/928/1/012018}}.

\bibitem{art:Muller}
F.~Müller, I.~Keshelashvili, D.~Mchedlishvili, {LYSO crystal testing for an
  EDM polarimeter}, J. Phys. Conf. Ser. 928~(1) (2017) 012019.
\newblock \href {https://doi.org/10.1088/1742-6596/928/1/012019}
  {\path{doi:10.1088/1742-6596/928/1/012019}}.

\bibitem{www:SensL}
SensL, "sensl - silicon photomultipliers \& spads",
  \url{http://sensl.com/downloads/ds/DS-MicroJseries.pdf} (2017).

\bibitem{www:Ti_regulator}
{Texas Instruments}, "precision voltage regulators",
  \url{http://www.ti.com/lit/ds/symlink/ua723.pdf} (2017).

\bibitem{www:Ti_dualAmp}
{Texas Instruments}, "lm158/lm258/lm358/lm2904 low power dual operational
  amplifiers", \url{http://www.ti.com/lit/ds/symlink/lm158-n.pdf} (2017).

\bibitem{www:MAX63}
{Maxim Integrated}, "max6325/41/50 ds",
  \url{https://datasheets.maximintegrated.com/en/ds/MAX6325-MAX6350.pdf}
  (2001).

\bibitem{www:Ti_prec_opamp}
{Texas Instruments}, "opax227 and opax228 high precision, low noise operational
  amplifiers,", \url{http://www.ti.com/lit/ds/symlink/opa4228.pdf} (2017).

\bibitem{www:Ferrara}
{University of Ferrara}, {Homepage}, \url{http://www.unife.it/international}.

\bibitem{www:RedPitaya}
{Red Pitaya}, "red pitaya", \url{https://www.redpitaya.com/index2} (2018).

\bibitem{www:altium}
{Altium Limited}, {"Homepage"}, \url{https://www.altium.com} (2020).

\bibitem{www:multisim}
{National Instruments Corporation}, {"Homepage"}, \url{https://www.ni.com}
  (2020).

\bibitem{www:CERN}
CERN, "root a data analysis framework", \url{https://root.cern.ch/}
  (2014-2018).

\bibitem{www:Keithley}
Tektronix, "keithley series 2700 multimeter/data acquisition/switch systems",
  \url{https://www.tek.com/keithley-switching-and-data-acquisition-systems/keithley-2700-multimeter-data-acquisition-switch-sys}
  (2018).

\bibitem{www:LM317}
{Texas Instruments}, "lm117hv/lm317hv 3-terminal adjustable regulator datasheet
  (rev. d)", \url{http://www.ti.com/lit/ds/symlink/lm317hv.pdf} (2017).

\bibitem{www:AD_mux}
{Analog Device}, "adg5206/adg5207 (rev. a)",
  \url{https://www.analog.com/media/en/technical-documentation/data-sheets/ADG5206_5207.pdf}
  (2012-2013).

\bibitem{www:VS_mux}
{Vishay Siliconix}, "8-ch/dual 4-ch high-performance cmos analog multiplexers",
  \url{https://www.vishay.com/docs/70062/dg408.pdf} (2017).

\bibitem{www:port_expander}
Microchip, "16-bit i/o expander with open-drain outputs",
  \url{http://ww1.microchip.com/downloads/en/DeviceDoc/22103a.pdf} (2008).

\bibitem{thes:oj}
O.~Javakhishvili, Power supply development for jedi polarimetry, Master's
  thesis, Agricultural University of Georgia, Tbilisi, Georgia (2018).

\end{thebibliography}
%\printbibliography 

\end{document}